\documentclass[twocolumn,preprintnumbers,aps,prl,showpacs,amsmath,amssymb,superscriptaddress]{revtex4-1}

\usepackage{graphicx}
\usepackage{dcolumn}
\usepackage{color}
\usepackage{bm}
\usepackage{amsmath}
\usepackage{dcolumn}

\DeclareMathOperator{\arcsinh}{arcsinh}
\begin{document}

\title{Entangled coherent states by mixing squeezed vacuum and coherent light}
\author{Yonatan Israel}
\altaffiliation{Current address: Physics Department, Stanford University, 382 Via Pueblo Mall, Stanford, California 94305, USA}
\affiliation{%
Department of Physics of Complex Systems, Weizmann Institute of
Science, Rehovot 76100, Israel}
\author{Lior Cohen}
\affiliation{%
Racah Institute of Physics, Hebrew University of Jerusalem, Jerusalem 91904, Israel}
\author{Xin-Bing Song}
\affiliation{%
Department of Physics of Complex Systems, Weizmann Institute of
Science, Rehovot 76100, Israel}
\author{Jaewoo Joo}
\affiliation{School of Electronic and Electrical Engineering, University of Leeds, Leeds, LS2 9JT, UK}
\affiliation{Clarendon Laboratory, University of Oxford, Parks Road, Oxford OX1 3PU, UK}
\author{Hagai S. Eisenberg}
\affiliation{%
Racah Institute of Physics, Hebrew University of Jerusalem, Jerusalem 91904, Israel}
\author{Yaron Silberberg}%
\affiliation{%
Department of Physics of Complex Systems, Weizmann Institute of
Science, Rehovot 76100, Israel}

\date{\today}

\begin{abstract}
Entangled coherent states are shown to emerge, with high fidelity, when mixing coherent and squeezed vacuum states of light on a beam-splitter. These maximally entangled states, where photons bunch at the exit of a beam-splitter, are measured experimentally by Fock-state projections. Entanglement is examined theoretically using a Bell-type nonlocality test and compared with ideal entangled coherent states. We experimentally show nearly perfect similarity with entangled coherent states for an optimal ratio of coherent and squeezed vacuum light. In our scheme, entangled coherent states are generated deterministically with small amplitudes, which could be beneficial, for example, in deterministic distribution of entanglement over long distances.
\end{abstract}

\pacs{42.50.-p 42.50.Dv 42.50.St}

\maketitle

\section{I. Introduction}

Entanglement is a defining feature of quantum mechanics with important implications to fundamental concepts, as well as for applications. Quantum states of light that exhibit entanglement were extensively employed in tests of the foundations of quantum theory \cite{AspectEPR82,ComplementarityTest13,Foundations2015} and are essential in quantum computing, quantum communication and quantum metrology \cite{QuantumMetrologyReview2011,WalmsleyReview15}. An intriguing class of states are the entangled coherent states (ECS), which contain coherent states (CS) $|\alpha\rangle$, in an equal superposition of being in either one of two possible paths \cite{ECS_Sanders1992,Gerry_ECSinequlaities,ECS_review_Sanders_2012}:
\begin{equation}\label{eq:ECS}
    |\psi_{ECS}^{\alpha}\rangle = \mathcal{N}_{\alpha} \left( |\alpha,0\rangle + |0,\alpha\rangle   \right),
\end{equation}
where $\mathcal{N}_{\alpha} = 1/\sqrt{2(1+e^{-\left|\alpha\right|^2})}$ is a normalization factor.

ECS manifest entanglement of coherent states - the most classical physical states, and are therefore fundamentally intriguing as they describe coherent states that are entangled with the vacuum (Eq. \ref{eq:ECS}). 
These states are also potentially useful in various applications of quantum technology.
It has been suggested that ECS could be advantageous resources for quantum information processing and quantum metrology \cite{ECS_review_Sanders_2012}, showing high tolerance against lossy quantum channels and interferometers \cite{ECS_loss_QI_PRA2010,PhysRevLett.Joo}, as well as reaching the Heisenberg limit in interferometry.

To create ECS, it has been suggested to make use of other non-classical Schr\"{o}dinger cat-states known as (even) coherent state superpositions (CSS) \cite{CSStoECS_Luis_PRA2001,ECS_review_Sanders_2012}:
\begin{align}\label{eq:CSS}
    |\psi_{CSS}^\beta \rangle &= \tilde{\mathcal{N}}_{\beta}\left(|\beta\rangle +  |-\beta\rangle\right)
\end{align}
where $\tilde{\mathcal{N}}_\beta = 1/\sqrt{2(1+e^{-2\left|\beta\right|^2})}$. Experimental realizations of CSS \cite{CSS_2phSub_PRL08,CSS_3phSub_PRA10} and then of ECS \cite{ECS_Grangier_NPhys2009} have mainly relied on a non deterministic technique involving photon subtraction, a probabilistic approach that is typically inefficient.
Deterministic schemes for generating CSS and ECS could be significantly more resource-effective. Such techniques using nonlinear interferometry were studied theoretically \cite{CSS_generation_Gerry_PRA99,NLI_GerryPRA02,ECS_EIT_PRA03}, but so far were not experimentally demonstrated \cite{CSS_review_08,ECS_review_Sanders_2012}.
In this work, we demonstrate experimentally a deterministic method for generating ECS by using deterministic squeezed vacuum (SV) and CS sources, and without resorting to probabilistic approaches in generating ECS, such as photon subtraction or post-selection \cite{ECS_Grangier_NPhys2009}.

ECS share similar properties with another class of entangled states, known as NOON states,
\begin{equation}\label{eq:NOON}
|\psi_{NOON}^N\rangle = (|N,0\rangle + |0,N\rangle)/\sqrt{2},
\end{equation}
where $N$ photons, rather than coherent states, are superposed in two modes. ECS are comprised of superpositions of NOON states \cite{Gerry_ECSinequlaities}, and both are capable of measurement sensitivities at the Heisenberg limit. While realizing NOON and ECS states with high intensities has been a long standing challenge \cite{CONTEMPPHYS08DOWLING,ECS_review_Sanders_2012}, since both states are prone to loss, ECS were proven to be more resilient in the context of quantum metrology \cite{PhysRevLett.Joo,QFI_Loss_ECS_NOON}.

Recently, it has been shown that mixing of coherent and squeezed vacuum light could give rise to NOON states \cite{HofmannPRA2007,Pezze2008PRL}, which were demonstrated up to $N=5$ \cite{AFEKSCIENCE,IsraelPRA2012,SteinbergNOONPRL2014}. In that approach, NOON states resulted from post-selecting $N$ photons after interfering SV and CS on a beam-splitter. In the current work we show theoretically and experimentally that the same system can be used to generate deterministically low-amplitude ECS with high fidelity.

\section{II. Theoretical analysis}
Fig. \ref{fig:beamsplitter}(a) illustrates schematically a process for preparing a perfect ECS $|\psi_{ECS}^{\alpha}\rangle$ 
by mixing a coherent state ($|\beta\rangle$
with a CSS $|\psi_{CSS}^{\beta}\rangle$
on a 50/50 beam splitter. Here $\beta = \alpha/\sqrt{2}$ \cite{CSStoECS_Luis_PRA2001}, and the average photon number in this state is
\begin{equation}\label{eq:n_avg_ECS}
\bar{n} = |\alpha|^2/\left(1+e^{-|\alpha|^2}\right).
\end{equation}
\subsection{A. Squeezed vacuum and coherent state interference}
\begin{figure}[tbh!]
  \centering
  \includegraphics[width=\columnwidth]{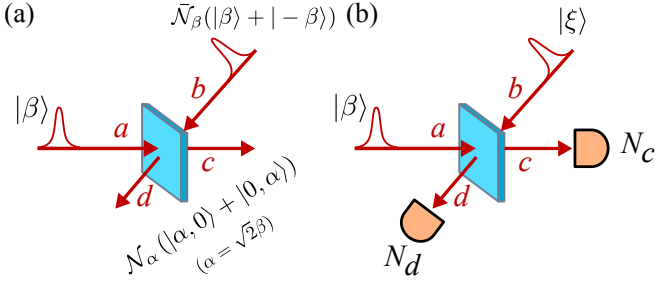}\\
\caption{(color online). Schematics for generating (a) ideal and (b) approximated entangled coherent states (ECS). A 50/50 beam splitter combines a coherent state $|\beta\rangle$ at port $a$ and (a) coherent state superpositions (CSS) $\tilde{\mathcal{N}}_{\beta}\left(|\beta\rangle +  |-\beta\rangle\right)$ at port $b$ to result with an exact ECS $\mathcal{N}_{\alpha} \left( |\alpha,0\rangle + |0,\alpha\rangle   \right)$, in ports $c,d$. (b) When the squeezed vacuum state $|\xi\rangle$ enters port $b$ instead of CSS, the result in ports $c,d$ approximates ECS (see text). In our experiment, a joint photon number measurement $N_c,N_d$ is performed at modes $c,d$, respectively, using photon number-resolving detectors.
 }\label{fig:beamsplitter}
\end{figure}

Consider now a similar system, where a CS $|\beta\rangle_a$ is mixed with a SV state $|\xi\rangle_b$ on a $50\%$ beam-splitter, as shown in Fig. \ref{fig:beamsplitter}(b). These input states can be defined in Fock basis \cite{IntroQuantOPt} as
\begin{align}
&|\beta\rangle = e^{-|\beta|^2/2}\sum_{n=0}^{\infty}\cfrac{\beta^n}{\sqrt{n!}}|n\rangle, \qquad\qquad\qquad \beta=|\beta|e^{i\phi},\label{eq:CS}\\
&|\xi\rangle = \cfrac{1}{\sqrt{\cosh r}} \sum_{m=0}^{\infty} (-1)^m \cfrac{\sqrt{(2m)!}}{2^m m!} (e^{i\theta}\tanh r)^m|2m\rangle \label{eq:SV},
\end{align}
where the phase of $|\beta\rangle$ and $|\xi\rangle$ are $\phi$ and $\theta$, respectively, while the relative phase of the two input states is $\theta-\phi$. The state at the output of the BS in modes $c$ and $d$ (Fig. \ref{fig:beamsplitter}(b)) is denoted by $|\psi_{out}\rangle$. The probability of $|\psi_{out}\rangle_{c,d}$ for $N_c$ and $N_d$ photons simultaneously at the output of the beam-splitter is given by \cite{JooECSWorkshop}:
\begin{align}\label{eq:Mixed_SV_CS}
P_{N_c,N_d}(\beta,r,\theta) = |\langle N_c,N_d | \psi_{out}\rangle_{c,d}|^2.
\end{align}

\subsection{B. Fidelity of ECS with mixed CS and SV}
Now, we show that the state $|\psi_{out}\rangle_{c,d}$ that is obtained when we mix a coherent state not with the ideal CSS state, but rather with a SV state, is still a  good approximation of the ECS. It should be noted that both SV (Eq. \ref{eq:SV}) and CSS (Eq. \ref{eq:CSS}) are composed of only even photon numbers and can be made approximately similar \cite{SV_approx_kittens_JOptB03}. This similarity can be evaluated through the fidelity between the two states \cite{JozsaFidelity}:
\begin{align}\label{eq:fidelity}
    F &= |\langle \psi_{ECS}^{\alpha} | \psi_{out} \rangle|^2 = |\langle \psi_{CSS} ^{\alpha/\sqrt{2}} | \xi \rangle|^2 = \nonumber\\ 
&=\frac{1}{\cosh r \cosh\big(\frac{|\alpha|^2}{2}\big)} e^{-\frac{|\alpha|^2}{2}\left(\cos(\theta-2\phi)\tanh r\right)}.
\end{align}
%
An optimal value of this fidelity (Eq. \ref{eq:fidelity}) is achieved for the following parameter relation:
\begin{align}
&r = \arcsinh(|\alpha|^2)/2,\nonumber\\
&\theta = 2\phi + \pi,\label{eq:max_fidelity}
\end{align}
where Eq. \ref{eq:max_fidelity} conditions the amplitude and phase of the SV to that of the CSS (see also Supplemental material \cite{SUPPLEMENTARY}). 


\begin{figure}[t]
 \centering
 \includegraphics[width=\columnwidth]{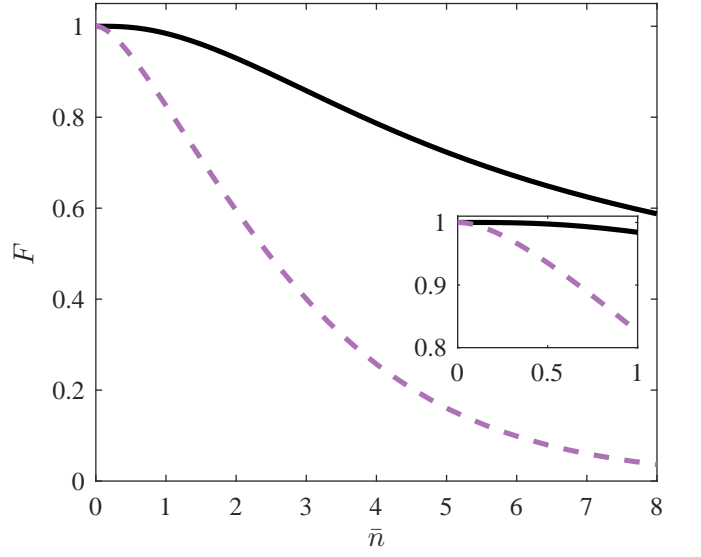}\\
\caption{(color online). The fidelity between ECS and states generated by mixing CS and SV, $F = |\langle \psi_{ECS}^{\alpha} | \psi_{out} \rangle|^2$, for the optimal SV amplitude as a function of the total photon number on average $\bar{n}$ (Eq. \ref{eq:n_avg_ECS}) in solid black. The inset shows $F$ for low values of the average photon number. The fidelity between CSS and the vacuum state ($|vac\rangle=|0\rangle$) is presented for comparison in dashed purple.}\label{fig:fidelity_CSS_SV_eq_navg}
\end{figure}

In Fig. \ref{fig:fidelity_CSS_SV_eq_navg}, the solid line presents the fidelity (Eq. \ref{eq:fidelity}) for the optimal values of SV (Eq. \ref{eq:max_fidelity}), showing that indeed, nearly perfect low amplitude ECS can be achieved using CS and SV, i.e. $F\approx1$ for $\bar{n},< 1$.
We note that the criteria in Eq. \ref{eq:max_fidelity} for the weak amplitudes regime ($\alpha,r\ll1$) coincides with the condition of setting the number of photon pairs of CS and SV to be equal, as needed for generating NOON states \cite{HofmannPRA2007,Pezze2008PRL,AFEKSCIENCE}.
The fidelity between CSS and the vacuum state (Fig. \ref{fig:fidelity_CSS_SV_eq_navg}, dashed line) is shown for comparison; this fidelity corresponds to the case of replacing the SV (Fig. \ref{fig:beamsplitter}(b)) with the vacuum state (Eq. \ref{eq:fidelity}), while the CS remains the other input to the BS. Note that the fidelity in this classical case is lower than the fidelity between ECS and the states generated by mixing CS and SV for all average photon numbers (see Fig. \ref{fig:fidelity_CSS_SV_eq_navg}).
Although the size of the ECS amplitude is relatively small, it can still violate the Bell inequality, as will be shown next.

\subsection{C. Nonlocality and the Janssens inequality}
A unique quantum property of ECS relates to its nonlocal correlations whereby multiple particles are all in one mode or the other. Such nonlocal properties are typically examined through the violation of Bell inequalities \cite{IntroQuantOPt}. ECS were previously shown to violate several types of such Bell-type inequalities, including a modified version of the Janssens inequalities \cite{JANSSENS,WildfeuerPRA07} that uses measurements of phase-space operators \cite{Gerry_ECSinequlaities}. 
We show here theoretically that the approximate ECS that result from mixing a CS and SV violates the inequalities as well. Measuring these inequalities in experiment requires homodyne detection, and is not accessible with our current detection setup.

We recall the expectation values of single- and two-mode phase-space operators on the modes $c$ and $d$:
\begin{align}
&Q_{c}(\mu) = \langle \psi_{out}| \hat Q_{c}(\mu) \otimes \hat I_{d} |\psi_{out}\rangle, \label{eq:Q_c}\\
&Q_{d}(\nu) = \langle \psi_{out}| \hat I_{c} \otimes \hat Q_{d}(\nu) |\psi_{out}\rangle, \label{eq:Q_d}\\
&Q_{c,d}(\mu,\nu) = \langle \psi_{out}| \hat Q_{c}(\mu) \otimes \hat Q_{d}(\nu) |\psi_{out} \label{eq:Q_cd}\rangle.
\end{align}
\begin{figure}[t]
 \centering
 \includegraphics[width=\columnwidth]{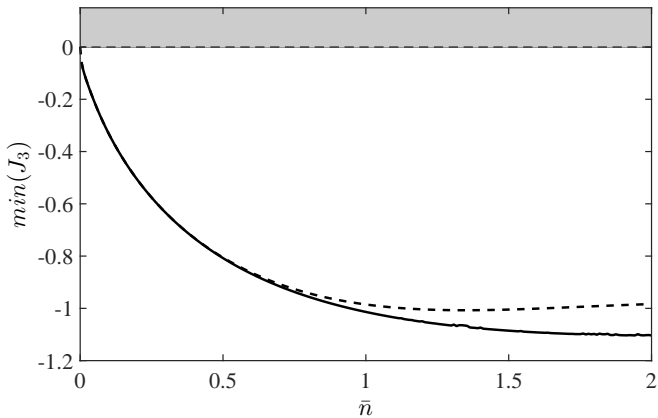}
\caption{(color online). Violation of the third Janssen inequality $J_3\leq0$ is shown below the gray shaded area, for the state $|\psi\rangle_{out}$ produced by mixing CS and SV with the optimal parameters of Eq. \ref{eq:max_fidelity} (solid line), and for an ideal ECS (dashed line), as a function of the total average photon number  $\bar{n}$. }\label{fig:J_3_nbar}
\end{figure}
Here $\hat Q_{j}(\mu)=|\mu\rangle\langle\mu|$ is a projection operator of the state in mode $j$ onto a coherent state of complex amplitude $\mu$, and $\hat I_{j}$ is the unity operator acting on mode $j$ (for $j=c,d$). Following Ref. \cite{WildfeuerPRA07}, a modified version of the third Janssens inequality can then be written as
\begin{align}\label{eq:J3}
J_3 = \,& Q(\alpha) - Q(\alpha,\beta) - Q(\alpha,\gamma) - Q(\alpha,\delta)\nonumber\\
 &+ Q(\beta,\gamma) + Q(\beta,\delta) + Q(\gamma,\delta)\leq0,
\end{align}
where $\alpha,\beta,\gamma$ and $\delta$ are any complex number. In Eq. \ref{eq:J3} the mode indices are omitted, meaning the $Q(\alpha)$ can be measured in either mode $c$ or $d$, and the joint two-mode expectation values are always measured between modes $c$ and $d$.

A minimizing procedure carried out on the parameters $\alpha,\beta,\gamma,\delta$ leads to a violation of the inequality $J_3\leq0$; this is shown in Fig. \ref{fig:J_3_nbar} for any given average photon number of the states generated in our scheme, $|\psi_{out}\rangle$, as well as for ECS, following a $\pi/2$ phase shift in mode $d$ (Fig. \ref{fig:beamsplitter}, see also Ref. \cite{Gerry_ECSinequlaities}). It is shown that the minimal value of $J_3$ merges for both states for low amplitudes, and deviates for larger average photon numbers, starting at $\bar n \approx 1$. We note that a similar analysis was recently done for CS mixed with photon subtracted SV \cite{toppel2016CV_subtSV}.

\section{III. Experimental setup and results}
\begin{figure}[t]
\centering
\includegraphics[width=\columnwidth]{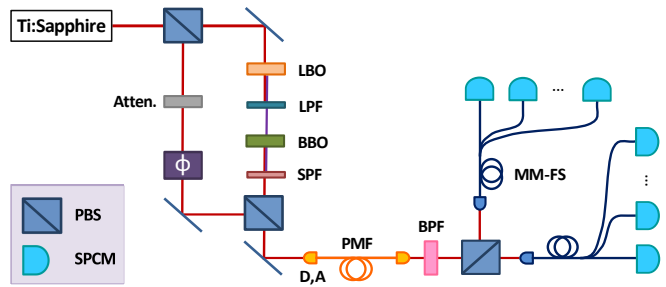}\\
\caption{(color online). Experimental setup for generation of entangled coherent states.
Detailed layout of the setup. 120-fs pulses from a Ti:sapphire oscillator
operated at 80 MHz are up-converted using a lithium triborate
(LBO) crystal, short pass filtered, and then down-converted using
a beta barium borate (BBO) crystal, generating a squeezed vacuum state, having correlated photon
pairs at the original wavelength (808 nm). This squeezed vacuum ($H$ polarization)
is mixed with attenuated coherent light ($V$ polarization)
on a polarizing beam-splitter (PBS). A thermally induced drift in the
relative phase is corrected every few minutes with the use of a liquid
crystal phase retarder, $\phi$. The spatial and
spectral modes are matched using a polarization-maintaining fiber
(PMF) and a 3-nm (full width at half max) bandpass filter (BPF).
CS and SV are mixed by a 50/50 beam-splitter transformation (Fig. \ref{fig:beamsplitter}(b)) in a collinear, polarization-based
inherently phase-stable design, by using a PMF fiber aligned at $\pm45^{\circ}$ ($D,A$) polarization axes, where ECS are realized.
Photon-number resolving detection is performed using an array of 16 single-photon
counting modules (SPCM, Perkin Elmer), and 1:8 multi-mode fiber splitters (MM-FS).}\label{fig:setup}
\end{figure}

Since an ECS is a superposition of NOON states for every photon number, we will show that the photon number distribution forms a corner distribution, that is, $P_{N_c,N_d} = |C_{N_c,N_d}|^2 $ (Eq. \ref{eq:Mixed_SV_CS}) is approximately $P_{N_c,N_d} = 0$ for $N_c\neq0$ and $N_d\neq0$.
The experimental setup (Fig. \ref{fig:setup}) is similar to the one used for the generation of NOON states  \cite{AFEKSCIENCE,IsraelPRA2012,NOONMicroscopyPRL2014}. Squeezed vacuum is produced via spontaneous parametric down-conversion (SPDC) and is mixed with a coherent state with indistinguishable spatial and spectral modes. These two sources are prepared in two orthogonal polarization modes ($H$ and $V$) and are combined by a polarizing beam-splitter (PBS). A polarization-maintaining fiber with axes oriented at $\pm45^{\circ}$ ($D,A$) is used to implement the BS of Fig. \ref{fig:beamsplitter}(b) in a collinear geometry. A second PBS sends the photons in each polarization mode to two photon-number resolving detectors based each on an 1:8 fiber splitter and 8 single-photon avalanche photon detectors, to record $N_c$ and $N_d$.

\begin{figure*}[t!] 
  \centering
  \includegraphics[width=0.9\textwidth]{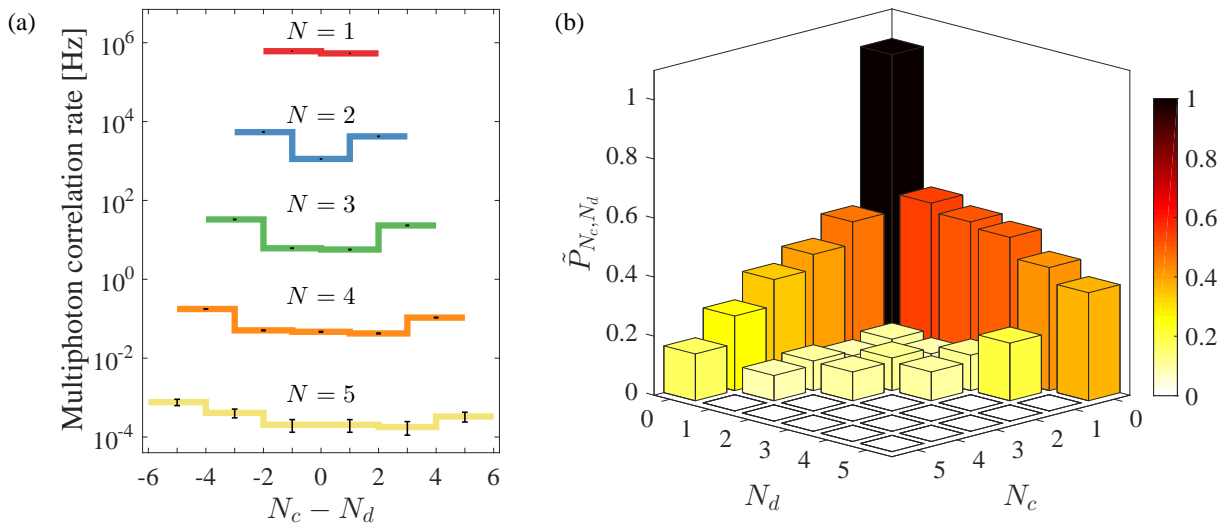}
  \caption{(color online). Experimental Fock projection measurements of coherent and squeezed vacuum light interfered on a 50:50 beam-splitter. (a) $N$-photon correlation rates plotted against the photon number difference between the output ports of the beam-splitter, $N_c-N_d$. Error bars represent the statistical standard error of the 24 hour long measurement. (b) Multiphoton correlation probabilities, normalized for every number of measured photons, $\tilde{P}_{N_c,N_d} = P_{N_c,N_d}/\left(\sum_{k=0}^{N}P_{k,N-k}\right)$.
  \label{fig:ExpRes}}
\end{figure*}

The results of the measured $P_{N_c,N_d}$ are presented in Fig. \ref{fig:ExpRes}. It is clear from these measurements that for any number of photons coming out of the beam, photons are highly bunched, that is, most are going to either port $c$ or port $d$. As shown in Fig. \ref{fig:ExpRes}(b), the probability for a photon correlation, normalized for every number of measured photons (see caption, Fig. \ref{fig:ExpRes}), is higher on the corner of the plot. An ideal ECS should have vanishing probability for all intermediate photon distributions.

%
%

\begin{figure}[htb!]
 \centering 
 \includegraphics[width=\columnwidth]{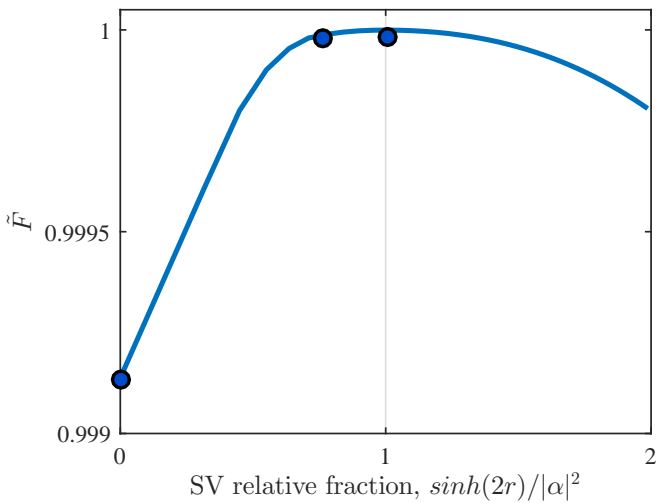}\\
\caption{(color online). The similarity $\tilde{F}$ between the state generated by mixing of SV with CS and ECS in our setup, accounting for loss ($\eta = 0.1$, see text), for various amounts of SV, using experimental (circles) and simulated (solid line) photon correlation measurements. The approximate ECS is achieved for the optimal SV fraction of $\sinh(2r)/|\alpha|^2 = 1$ (Eq \ref{eq:max_fidelity}), showing maximal similarity to ECS.}\label{fig:similarity}
\end{figure}
In order to quantify the similarity of the multi-photon correlation measurements $P_{m,n}$ between the approximate states and perfect ECS, we calculated $\tilde{F} = (\sum_{m,n} \sqrt{P_{m,n} \cdot P^{ECS}_{m,n}})^2$ \cite{QuantumWalks2010}, where $P^{ECS}_{m,n}$ is the photon number correlation for ECS. Fig. \ref{fig:similarity} presents the similarity $\tilde{F}$ obtained by varying the amount of squeezed vacuum in the experimental measurements (cirles) or simulation (solid line). Here, $P^{ECS}_{m,n}$ was calculated from a simulation for perfect generation of an ECS (Eq. \ref{eq:ECS}) as would be detected in our setup, accounting for our detection scheme and losses ($\eta = 0.1$), using no fit parameters \cite{AFEKSCIENCE,IsraelPRA2012}. The CS amplitude was decreased from $\beta=0.75$ to $\beta = 0.45$, as the fraction of SV is increased. These results show that a maximal similarity is achieved for its optimal parameters of Eq. \ref{eq:max_fidelity}. We note that the experimental state has non-zero off-corner terms as can be seen in Fig. 5. These terms reduce the similarity to an ECS, but their contribution is exponentially small due to their exponentially small probabilities compared with the probabilities on the corner.

\section{IV. Discussion}
The measurements presented in Fig. \ref{fig:ExpRes} could have appeared to result from a mixed state. To show that this is not the case, and rather that these states are in fact ECS, we have derived a measure for the purity of ECS in our scheme. Our measurements suggest that states in our setup are indeed close to pure ECS (see Supplemental Material \cite{SUPPLEMENTARY}).

Our method provides a simple and deterministic route to generate entangled coherent states. We achieve amplitude values that are comparable with previous experimental realizations, while these relied on a photon-subtraction technique, resulting in an indeterministic photon source \cite{ECS_Grangier_NPhys2009}.
ECS have also been realized recently in super-conducting circuits deterministically \cite{YaleECS2016}, however translating these states to traveling waves, as typically required in application of quantum metrology and quantum communications, has not been demonstrated yet.

The average photon number of the state in our experimental realization was $\bar{n}=0.15$, for which the fidelity to ECS is theoretically $F\approx1$ (see Fig. \ref{fig:fidelity_CSS_SV_eq_navg}).
Previous measurements of quantum state tomography in $N$-photon subspaces of the states generated in our setup have shown high fidelities to NOON states \cite{IsraelPRA2012}, in agreement with ECS (Eq. \ref{eq:ECS}).
Scaling up our approach to higher average photon numbers is therefore highly desired, where high fidelity ECS with $\bar n \sim 1$ should be achievable, using a more energetic source of SV (for example, Ref. \cite{CSS_3phSub_PRA10}).
It should be noted that even low amplitude could prove to be beneficial in some applications; for example, it was recently shown that there is an advantage in using low amplitude ECS ($\alpha\approx 1$) over larger ones for quantum communication \cite{ECS-Qcomm_sangouard2010}.

Imperfections in the experiment had two main causes. One reason involved the distinguishability between two independent photons from the CS and SV state, which limits their interference visibility (two-photon interference visibility is $v=0.91\pm 0.02$ with 95\% confidence level  \cite{SUPPLEMENTARY}) and mainly increases the probability for $P_{N_c,N_d}$ for $N_c, N_d\neq 0$. The other reason has to do with the fiber splitters. Coupling efficiency difference of about $12.8\%$ as well as non uniform splitting ratios result in skewness of the plots in Fig. \ref{fig:ExpRes}, towards port $d$.

\section{V. Summary}
In summary, we have shown that the interference of coherent light and squeezed light on a beam-splitter can generate low amplitude entangled coherent state with high fidelity. These states violate a Bell-type inequality, similarly to entangled coherent states. We have experimentally realized these states, and analyzed them through photon number detection, showing a pronounced corner-like two-mode distribution of photons, with maximal overlap for an optimal fraction of squeezed vacuum and coherent light. Our method benefits from a relatively simple setup that allows a deterministic route to generating entangled coherent states without resorting to inefficient photon subtraction. Such approach could become useful for long-distance entanglement distribution \cite{LongDistEntDist_PRL10}, particularly with low amplitudes \cite{ECS-Qcomm_sangouard2010} and using lossy channels \cite{CV_LossyEntSwappingPRA16}.

\begin{acknowledgments}
This work was supported by DIP - German-Israeli Project Cooperation, by the BSF-NSF grant \#2014719, by Icore - Israeli Centre of Research Excellence “Circle of Light”, and by the Crown Photonics Center.
Correspondence and requests for materials should be addressed to Y. I. ~(email: yisrael@stanford.edu).
\end{acknowledgments}

\bibliographystyle{apsrev}

\begin{thebibliography}{38}
\expandafter\ifx\csname natexlab\endcsname\relax\def\natexlab#1{#1}\fi
\expandafter\ifx\csname bibnamefont\endcsname\relax
  \def\bibnamefont#1{#1}\fi
\expandafter\ifx\csname bibfnamefont\endcsname\relax
  \def\bibfnamefont#1{#1}\fi
\expandafter\ifx\csname citenamefont\endcsname\relax
  \def\citenamefont#1{#1}\fi
\expandafter\ifx\csname url\endcsname\relax
  \def\url#1{\texttt{#1}}\fi
\expandafter\ifx\csname urlprefix\endcsname\relax\def\urlprefix{URL }\fi
\providecommand{\bibinfo}[2]{#2}
\providecommand{\eprint}[2][]{\url{#2}}

\bibitem[{\citenamefont{Aspect et~al.}(1982)\citenamefont{Aspect, Grangier, and
  Roger}}]{AspectEPR82}
\bibinfo{author}{\bibfnamefont{A.}~\bibnamefont{Aspect}},
  \bibinfo{author}{\bibfnamefont{P.}~\bibnamefont{Grangier}}, \bibnamefont{and}
  \bibinfo{author}{\bibfnamefont{G.}~\bibnamefont{Roger}},
  \bibinfo{journal}{Physical Review Letters} \textbf{\bibinfo{volume}{49}},
  \bibinfo{pages}{91} (\bibinfo{year}{1982}).

\bibitem[{\citenamefont{Weston et~al.}(2013)\citenamefont{Weston, Hall,
  Palsson, Wiseman, and Pryde}}]{ComplementarityTest13}
\bibinfo{author}{\bibfnamefont{M.~M.} \bibnamefont{Weston}},
  \bibinfo{author}{\bibfnamefont{M.~J.~W.} \bibnamefont{Hall}},
  \bibinfo{author}{\bibfnamefont{M.~S.} \bibnamefont{Palsson}},
  \bibinfo{author}{\bibfnamefont{H.~M.} \bibnamefont{Wiseman}},
  \bibnamefont{and} \bibinfo{author}{\bibfnamefont{G.~J.} \bibnamefont{Pryde}},
  \bibinfo{journal}{Phys. Rev. Lett.} \textbf{\bibinfo{volume}{110}},
  \bibinfo{pages}{220402} (\bibinfo{year}{2013}).

\bibitem[{\citenamefont{Shadbolt et~al.}(2014)\citenamefont{Shadbolt, Mathews,
  Laing, and O'Brien}}]{Foundations2015}
\bibinfo{author}{\bibfnamefont{P.}~\bibnamefont{Shadbolt}},
  \bibinfo{author}{\bibfnamefont{J.~C.~F.} \bibnamefont{Mathews}},
  \bibinfo{author}{\bibfnamefont{A.}~\bibnamefont{Laing}}, \bibnamefont{and}
  \bibinfo{author}{\bibfnamefont{J.~L.} \bibnamefont{O'Brien}},
  \bibinfo{journal}{Nature Physics} \textbf{\bibinfo{volume}{10}},
  \bibinfo{pages}{278} (\bibinfo{year}{2014}).

\bibitem[{\citenamefont{Giovannetti et~al.}(2011)\citenamefont{Giovannetti,
  Lloyd, and Maccone}}]{QuantumMetrologyReview2011}
\bibinfo{author}{\bibfnamefont{V.}~\bibnamefont{Giovannetti}},
  \bibinfo{author}{\bibfnamefont{S.}~\bibnamefont{Lloyd}}, \bibnamefont{and}
  \bibinfo{author}{\bibfnamefont{L.}~\bibnamefont{Maccone}},
  \bibinfo{journal}{Nature Photonics} \textbf{\bibinfo{volume}{5}},
  \bibinfo{pages}{222} (\bibinfo{year}{2011}).

\bibitem[{\citenamefont{Walmsley}(2015)}]{WalmsleyReview15}
\bibinfo{author}{\bibfnamefont{I.~A.} \bibnamefont{Walmsley}},
  \bibinfo{journal}{Science} \textbf{\bibinfo{volume}{348}},
  \bibinfo{pages}{525} (\bibinfo{year}{2015}).

\bibitem[{\citenamefont{Sanders}(1992)}]{ECS_Sanders1992}
\bibinfo{author}{\bibfnamefont{B.~C.} \bibnamefont{Sanders}},
  \bibinfo{journal}{Phys. Rev. A} \textbf{\bibinfo{volume}{45}},
  \bibinfo{pages}{6811} (\bibinfo{year}{1992}).

\bibitem[{\citenamefont{Gerry et~al.}(2009)\citenamefont{Gerry, Mimih, and
  Benmoussa}}]{Gerry_ECSinequlaities}
\bibinfo{author}{\bibfnamefont{C.~C.} \bibnamefont{Gerry}},
  \bibinfo{author}{\bibfnamefont{J.}~\bibnamefont{Mimih}}, \bibnamefont{and}
  \bibinfo{author}{\bibfnamefont{A.}~\bibnamefont{Benmoussa}},
  \bibinfo{journal}{Phys. Rev. A} \textbf{\bibinfo{volume}{80}},
  \bibinfo{pages}{022111} (\bibinfo{year}{2009}).

\bibitem[{\citenamefont{Sanders}(2012)}]{ECS_review_Sanders_2012}
\bibinfo{author}{\bibfnamefont{B.~C.} \bibnamefont{Sanders}},
  \bibinfo{journal}{Journal of Physics A: Mathematical and Theoretical}
  \textbf{\bibinfo{volume}{45}}, \bibinfo{pages}{244002}
  (\bibinfo{year}{2012}).

\bibitem[{\citenamefont{Park and Jeong}(2010)}]{ECS_loss_QI_PRA2010}
\bibinfo{author}{\bibfnamefont{K.}~\bibnamefont{Park}} \bibnamefont{and}
  \bibinfo{author}{\bibfnamefont{H.}~\bibnamefont{Jeong}},
  \bibinfo{journal}{Physical Review A} \textbf{\bibinfo{volume}{82}},
  \bibinfo{pages}{062325} (\bibinfo{year}{2010}).

\bibitem[{\citenamefont{Joo et~al.}(2011)\citenamefont{Joo, Munro, and
  Spiller}}]{PhysRevLett.Joo}
\bibinfo{author}{\bibfnamefont{J.}~\bibnamefont{Joo}},
  \bibinfo{author}{\bibfnamefont{W.~J.} \bibnamefont{Munro}}, \bibnamefont{and}
  \bibinfo{author}{\bibfnamefont{T.~P.} \bibnamefont{Spiller}},
  \bibinfo{journal}{Phys. Rev. Lett.} \textbf{\bibinfo{volume}{107}},
  \bibinfo{pages}{083601} (\bibinfo{year}{2011}).

\bibitem[{\citenamefont{Luis}(2001)}]{CSStoECS_Luis_PRA2001}
\bibinfo{author}{\bibfnamefont{A.}~\bibnamefont{Luis}}, \bibinfo{journal}{Phys.
  Rev. A} \textbf{\bibinfo{volume}{64}}, \bibinfo{pages}{054102}
  (\bibinfo{year}{2001}).

\bibitem[{\citenamefont{Takahashi et~al.}(2008)\citenamefont{Takahashi, Wakui,
  Suzuki, Takeoka, Hayasaka, Furusawa, and Sasaki}}]{CSS_2phSub_PRL08}
\bibinfo{author}{\bibfnamefont{H.}~\bibnamefont{Takahashi}},
  \bibinfo{author}{\bibfnamefont{K.}~\bibnamefont{Wakui}},
  \bibinfo{author}{\bibfnamefont{S.}~\bibnamefont{Suzuki}},
  \bibinfo{author}{\bibfnamefont{M.}~\bibnamefont{Takeoka}},
  \bibinfo{author}{\bibfnamefont{K.}~\bibnamefont{Hayasaka}},
  \bibinfo{author}{\bibfnamefont{A.}~\bibnamefont{Furusawa}}, \bibnamefont{and}
  \bibinfo{author}{\bibfnamefont{M.}~\bibnamefont{Sasaki}},
  \bibinfo{journal}{Phys. Rev. Lett.} \textbf{\bibinfo{volume}{101}},
  \bibinfo{pages}{233605} (\bibinfo{year}{2008}).

\bibitem[{\citenamefont{Gerrits et~al.}(2010)\citenamefont{Gerrits, Glancy,
  Clement, Calkins, Lita, Miller, Migdall, Nam, Mirin, and
  Knill}}]{CSS_3phSub_PRA10}
\bibinfo{author}{\bibfnamefont{T.}~\bibnamefont{Gerrits}},
  \bibinfo{author}{\bibfnamefont{S.}~\bibnamefont{Glancy}},
  \bibinfo{author}{\bibfnamefont{T.~S.} \bibnamefont{Clement}},
  \bibinfo{author}{\bibfnamefont{B.}~\bibnamefont{Calkins}},
  \bibinfo{author}{\bibfnamefont{A.~E.} \bibnamefont{Lita}},
  \bibinfo{author}{\bibfnamefont{A.~J.} \bibnamefont{Miller}},
  \bibinfo{author}{\bibfnamefont{A.~L.} \bibnamefont{Migdall}},
  \bibinfo{author}{\bibfnamefont{S.~W.} \bibnamefont{Nam}},
  \bibinfo{author}{\bibfnamefont{R.~P.} \bibnamefont{Mirin}}, \bibnamefont{and}
  \bibinfo{author}{\bibfnamefont{E.}~\bibnamefont{Knill}},
  \bibinfo{journal}{Phys. Rev. A} \textbf{\bibinfo{volume}{82}},
  \bibinfo{pages}{031802} (\bibinfo{year}{2010}).

\bibitem[{\citenamefont{Ourjoumtsev et~al.}(2009)\citenamefont{Ourjoumtsev,
  Ferreyrol, Tualle-Brouri, and Grangier}}]{ECS_Grangier_NPhys2009}
\bibinfo{author}{\bibfnamefont{A.}~\bibnamefont{Ourjoumtsev}},
  \bibinfo{author}{\bibfnamefont{F.}~\bibnamefont{Ferreyrol}},
  \bibinfo{author}{\bibfnamefont{R.}~\bibnamefont{Tualle-Brouri}},
  \bibnamefont{and} \bibinfo{author}{\bibfnamefont{P.}~\bibnamefont{Grangier}},
  \bibinfo{journal}{Nature Physics} \textbf{\bibinfo{volume}{5}},
  \bibinfo{pages}{189} (\bibinfo{year}{2009}).

\bibitem[{\citenamefont{Gerry}(1999)}]{CSS_generation_Gerry_PRA99}
\bibinfo{author}{\bibfnamefont{C.~C.} \bibnamefont{Gerry}},
  \bibinfo{journal}{Phys. Rev. A} \textbf{\bibinfo{volume}{59}},
  \bibinfo{pages}{4095} (\bibinfo{year}{1999}).

\bibitem[{\citenamefont{Gerry et~al.}(2002)\citenamefont{Gerry, Benmoussa, and
  Campos}}]{NLI_GerryPRA02}
\bibinfo{author}{\bibfnamefont{C.~C.} \bibnamefont{Gerry}},
  \bibinfo{author}{\bibfnamefont{A.}~\bibnamefont{Benmoussa}},
  \bibnamefont{and} \bibinfo{author}{\bibfnamefont{R.~A.}
  \bibnamefont{Campos}}, \bibinfo{journal}{Phys. Rev. A}
  \textbf{\bibinfo{volume}{66}}, \bibinfo{pages}{013804}
  (\bibinfo{year}{2002}).

\bibitem[{\citenamefont{Paternostro et~al.}(2003)\citenamefont{Paternostro,
  Kim, and Ham}}]{ECS_EIT_PRA03}
\bibinfo{author}{\bibfnamefont{M.}~\bibnamefont{Paternostro}},
  \bibinfo{author}{\bibfnamefont{M.~S.} \bibnamefont{Kim}}, \bibnamefont{and}
  \bibinfo{author}{\bibfnamefont{B.~S.} \bibnamefont{Ham}},
  \bibinfo{journal}{Phys. Rev. A} \textbf{\bibinfo{volume}{67}},
  \bibinfo{pages}{023811} (\bibinfo{year}{2003}).

\bibitem[{\citenamefont{Glancy and de~Vasconcelos}(2008)}]{CSS_review_08}
\bibinfo{author}{\bibfnamefont{S.}~\bibnamefont{Glancy}} \bibnamefont{and}
  \bibinfo{author}{\bibfnamefont{H.~M.} \bibnamefont{de~Vasconcelos}},
  \bibinfo{journal}{J. Opt. Soc. Am. B} \textbf{\bibinfo{volume}{25}},
  \bibinfo{pages}{712} (\bibinfo{year}{2008}).

\bibitem[{\citenamefont{Dowling}(2008)}]{CONTEMPPHYS08DOWLING}
\bibinfo{author}{\bibfnamefont{J.}~\bibnamefont{Dowling}},
  \bibinfo{journal}{Contemporary physics} \textbf{\bibinfo{volume}{49}},
  \bibinfo{pages}{125} (\bibinfo{year}{2008}).

\bibitem[{\citenamefont{Zhang et~al.}(2013)\citenamefont{Zhang, Li, Yang, and
  Jin}}]{QFI_Loss_ECS_NOON}
\bibinfo{author}{\bibfnamefont{Y.~M.} \bibnamefont{Zhang}},
  \bibinfo{author}{\bibfnamefont{X.~W.} \bibnamefont{Li}},
  \bibinfo{author}{\bibfnamefont{W.}~\bibnamefont{Yang}}, \bibnamefont{and}
  \bibinfo{author}{\bibfnamefont{G.~R.} \bibnamefont{Jin}},
  \bibinfo{journal}{Phys. Rev. A} \textbf{\bibinfo{volume}{88}},
  \bibinfo{pages}{043832} (\bibinfo{year}{2013}).

\bibitem[{\citenamefont{Hofmann and Ono}(2007)}]{HofmannPRA2007}
\bibinfo{author}{\bibfnamefont{H.~F.} \bibnamefont{Hofmann}} \bibnamefont{and}
  \bibinfo{author}{\bibfnamefont{T.}~\bibnamefont{Ono}},
  \bibinfo{journal}{Phys. Rev. A} \textbf{\bibinfo{volume}{76}},
  \bibinfo{pages}{031806} (\bibinfo{year}{2007}).

\bibitem[{\citenamefont{Pezz\'e and Smerzi}(2008)}]{Pezze2008PRL}
\bibinfo{author}{\bibfnamefont{L.}~\bibnamefont{Pezz\'e}} \bibnamefont{and}
  \bibinfo{author}{\bibfnamefont{A.}~\bibnamefont{Smerzi}},
  \bibinfo{journal}{Phys. Rev. Lett.} \textbf{\bibinfo{volume}{100}},
  \bibinfo{pages}{073601} (\bibinfo{year}{2008}).

\bibitem[{\citenamefont{Afek et~al.}(2010)\citenamefont{Afek, Ambar, and
  Silberberg}}]{AFEKSCIENCE}
\bibinfo{author}{\bibfnamefont{I.}~\bibnamefont{Afek}},
  \bibinfo{author}{\bibfnamefont{O.}~\bibnamefont{Ambar}}, \bibnamefont{and}
  \bibinfo{author}{\bibfnamefont{Y.}~\bibnamefont{Silberberg}},
  \bibinfo{journal}{Science} \textbf{\bibinfo{volume}{328}},
  \bibinfo{pages}{879} (\bibinfo{year}{2010}).

\bibitem[{\citenamefont{Israel et~al.}(2012)\citenamefont{Israel, Afek, Rosen,
  Ambar, and Silberberg}}]{IsraelPRA2012}
\bibinfo{author}{\bibfnamefont{Y.}~\bibnamefont{Israel}},
  \bibinfo{author}{\bibfnamefont{I.}~\bibnamefont{Afek}},
  \bibinfo{author}{\bibfnamefont{S.}~\bibnamefont{Rosen}},
  \bibinfo{author}{\bibfnamefont{O.}~\bibnamefont{Ambar}}, \bibnamefont{and}
  \bibinfo{author}{\bibfnamefont{Y.}~\bibnamefont{Silberberg}},
  \bibinfo{journal}{Phys. Rev. A} \textbf{\bibinfo{volume}{85}},
  \bibinfo{pages}{022115} (\bibinfo{year}{2012}).

\bibitem[{\citenamefont{Rozema et~al.}(2014)\citenamefont{Rozema, Bateman,
  Mahler, Okamoto, Feizpour, Hayat, and Steinberg}}]{SteinbergNOONPRL2014}
\bibinfo{author}{\bibfnamefont{L.~A.} \bibnamefont{Rozema}},
  \bibinfo{author}{\bibfnamefont{J.~D.} \bibnamefont{Bateman}},
  \bibinfo{author}{\bibfnamefont{D.~H.} \bibnamefont{Mahler}},
  \bibinfo{author}{\bibfnamefont{R.}~\bibnamefont{Okamoto}},
  \bibinfo{author}{\bibfnamefont{A.}~\bibnamefont{Feizpour}},
  \bibinfo{author}{\bibfnamefont{A.}~\bibnamefont{Hayat}}, \bibnamefont{and}
  \bibinfo{author}{\bibfnamefont{A.~M.} \bibnamefont{Steinberg}},
  \bibinfo{journal}{Phys. Rev. Lett.} \textbf{\bibinfo{volume}{112}},
  \bibinfo{pages}{223602} (\bibinfo{year}{2014}).

\bibitem[{\citenamefont{Gerry and Knight}(2005)}]{IntroQuantOPt}
\bibinfo{author}{\bibfnamefont{C.~C.} \bibnamefont{Gerry}} \bibnamefont{and}
  \bibinfo{author}{\bibfnamefont{P.~L.} \bibnamefont{Knight}},
  \emph{\bibinfo{title}{Introductory Quantum Optics}}
  (\bibinfo{publisher}{Cambridge University Press}, \bibinfo{year}{2005}).

\bibitem[{\citenamefont{Joo et~al.}(2013)\citenamefont{Joo, Park, Jeong, Munro,
  Nemoto, and Spiller}}]{JooECSWorkshop}
\bibinfo{author}{\bibfnamefont{J.}~\bibnamefont{Joo}},
  \bibinfo{author}{\bibfnamefont{K.}~\bibnamefont{Park}},
  \bibinfo{author}{\bibfnamefont{H.}~\bibnamefont{Jeong}},
  \bibinfo{author}{\bibfnamefont{W.~J.} \bibnamefont{Munro}},
  \bibinfo{author}{\bibfnamefont{K.}~\bibnamefont{Nemoto}}, \bibnamefont{and}
  \bibinfo{author}{\bibfnamefont{T.~P.} \bibnamefont{Spiller}}, in
  \emph{\bibinfo{booktitle}{Proceedings of the First International Workshop on
  ECS and Its Application to QIS}} (\bibinfo{year}{2013}).

\bibitem[{\citenamefont{Gerry et~al.}(2003)\citenamefont{Gerry, Benmoussa, and
  Bruno}}]{SV_approx_kittens_JOptB03}
\bibinfo{author}{\bibfnamefont{C.~C.} \bibnamefont{Gerry}},
  \bibinfo{author}{\bibfnamefont{A.}~\bibnamefont{Benmoussa}},
  \bibnamefont{and} \bibinfo{author}{\bibfnamefont{K.~M.} \bibnamefont{Bruno}},
  \bibinfo{journal}{Journal of Optics B: Quantum and Semiclassical Optics}
  \textbf{\bibinfo{volume}{5}}, \bibinfo{pages}{109} (\bibinfo{year}{2003}).

\bibitem[{\citenamefont{Jozsa}(1994)}]{JozsaFidelity}
\bibinfo{author}{\bibfnamefont{R.}~\bibnamefont{Jozsa}},
  \bibinfo{journal}{Journal of Modern Optics} \textbf{\bibinfo{volume}{41}},
  \bibinfo{pages}{2315} (\bibinfo{year}{1994}).

\bibitem[{SUP()}]{SUPPLEMENTARY}
\bibinfo{note}{See Supplemental Material at [URL] for the purity of ECS and the weak amplitude description of SV and CS mixing in Fock basis.}

\bibitem[{\citenamefont{Janssens et~al.}(2004)\citenamefont{Janssens, Baets,
  and Meyer}}]{JANSSENS}
\bibinfo{author}{\bibfnamefont{S.}~\bibnamefont{Janssens}},
  \bibinfo{author}{\bibfnamefont{B.~D.} \bibnamefont{Baets}}, \bibnamefont{and}
  \bibinfo{author}{\bibfnamefont{H.~D.} \bibnamefont{Meyer}},
  \bibinfo{journal}{Fuzzy Sets and Systems} \textbf{\bibinfo{volume}{148}},
  \bibinfo{pages}{263 } (\bibinfo{year}{2004}), ISSN \bibinfo{issn}{0165-0114}.

\bibitem[{\citenamefont{Wildfeuer et~al.}(2007)\citenamefont{Wildfeuer, Lund,
  and Dowling}}]{WildfeuerPRA07}
\bibinfo{author}{\bibfnamefont{C.~F.} \bibnamefont{Wildfeuer}},
  \bibinfo{author}{\bibfnamefont{A.~P.} \bibnamefont{Lund}}, \bibnamefont{and}
  \bibinfo{author}{\bibfnamefont{J.~P.} \bibnamefont{Dowling}},
  \bibinfo{journal}{Phys. Rev. A} \textbf{\bibinfo{volume}{76}},
  \bibinfo{pages}{052101} (\bibinfo{year}{2007}).

\bibitem[{\citenamefont{T{\"o}ppel et~al.}(2016)\citenamefont{T{\"o}ppel,
  Chekhova, and Leuchs}}]{toppel2016CV_subtSV}
\bibinfo{author}{\bibfnamefont{F.}~\bibnamefont{T{\"o}ppel}},
  \bibinfo{author}{\bibfnamefont{M.~V.} \bibnamefont{Chekhova}},
  \bibnamefont{and} \bibinfo{author}{\bibfnamefont{G.}~\bibnamefont{Leuchs}},
  \bibinfo{journal}{arXiv preprint arXiv:1607.01296}  (\bibinfo{year}{2016}).

\bibitem[{\citenamefont{Israel et~al.}(2014)\citenamefont{Israel, Rosen, and
  Silberberg}}]{NOONMicroscopyPRL2014}
\bibinfo{author}{\bibfnamefont{Y.}~\bibnamefont{Israel}},
  \bibinfo{author}{\bibfnamefont{S.}~\bibnamefont{Rosen}}, \bibnamefont{and}
  \bibinfo{author}{\bibfnamefont{Y.}~\bibnamefont{Silberberg}},
  \bibinfo{journal}{Physical Review Letters} \textbf{\bibinfo{volume}{112}},
  \bibinfo{pages}{103604} (\bibinfo{year}{2014}).

\bibitem[{\citenamefont{Peruzzo et~al.}(2010)\citenamefont{Peruzzo, Lobino,
  Matthews, Matsuda, Politi, Poulios, Zhou, Lahini, Ismail, W{\"o}rhoff
  et~al.}}]{QuantumWalks2010}
\bibinfo{author}{\bibfnamefont{A.}~\bibnamefont{Peruzzo}},
  \bibinfo{author}{\bibfnamefont{M.}~\bibnamefont{Lobino}},
  \bibinfo{author}{\bibfnamefont{J.~C.} \bibnamefont{Matthews}},
  \bibinfo{author}{\bibfnamefont{N.}~\bibnamefont{Matsuda}},
  \bibinfo{author}{\bibfnamefont{A.}~\bibnamefont{Politi}},
  \bibinfo{author}{\bibfnamefont{K.}~\bibnamefont{Poulios}},
  \bibinfo{author}{\bibfnamefont{X.-Q.} \bibnamefont{Zhou}},
  \bibinfo{author}{\bibfnamefont{Y.}~\bibnamefont{Lahini}},
  \bibinfo{author}{\bibfnamefont{N.}~\bibnamefont{Ismail}},
  \bibinfo{author}{\bibfnamefont{K.}~\bibnamefont{W{\"o}rhoff}},
  \bibnamefont{et~al.}, \bibinfo{journal}{Science}
  \textbf{\bibinfo{volume}{329}}, \bibinfo{pages}{1500} (\bibinfo{year}{2010}).

\bibitem[{\citenamefont{Wang et~al.}(2016)\citenamefont{Wang, Gao, Reinhold,
  Heeres, Ofek, Chou, Axline, Reagor, Blumoff, Sliwa et~al.}}]{YaleECS2016}
\bibinfo{author}{\bibfnamefont{C.}~\bibnamefont{Wang}},
  \bibinfo{author}{\bibfnamefont{Y.~Y.} \bibnamefont{Gao}},
  \bibinfo{author}{\bibfnamefont{P.}~\bibnamefont{Reinhold}},
  \bibinfo{author}{\bibfnamefont{R.}~\bibnamefont{Heeres}},
  \bibinfo{author}{\bibfnamefont{N.}~\bibnamefont{Ofek}},
  \bibinfo{author}{\bibfnamefont{K.}~\bibnamefont{Chou}},
  \bibinfo{author}{\bibfnamefont{C.}~\bibnamefont{Axline}},
  \bibinfo{author}{\bibfnamefont{M.}~\bibnamefont{Reagor}},
  \bibinfo{author}{\bibfnamefont{J.}~\bibnamefont{Blumoff}},
  \bibinfo{author}{\bibfnamefont{K.}~\bibnamefont{Sliwa}},
  \bibnamefont{et~al.}, \bibinfo{journal}{Science}
  \textbf{\bibinfo{volume}{352}}, \bibinfo{pages}{1087} (\bibinfo{year}{2016}).

\bibitem[{\citenamefont{Sangouard et~al.}(2010)\citenamefont{Sangouard, Simon,
  Gisin, Laurat, Tualle-Brouri, and Grangier}}]{ECS-Qcomm_sangouard2010}
\bibinfo{author}{\bibfnamefont{N.}~\bibnamefont{Sangouard}},
  \bibinfo{author}{\bibfnamefont{C.}~\bibnamefont{Simon}},
  \bibinfo{author}{\bibfnamefont{N.}~\bibnamefont{Gisin}},
  \bibinfo{author}{\bibfnamefont{J.}~\bibnamefont{Laurat}},
  \bibinfo{author}{\bibfnamefont{R.}~\bibnamefont{Tualle-Brouri}},
  \bibnamefont{and} \bibinfo{author}{\bibfnamefont{P.}~\bibnamefont{Grangier}},
  \bibinfo{journal}{J. Opt. Soc. Am. B} \textbf{\bibinfo{volume}{27}},
  \bibinfo{pages}{A137} (\bibinfo{year}{2010}).

\bibitem[{\citenamefont{Brask et~al.}(2010)\citenamefont{Brask, Rigas, Polzik,
  Andersen, and S\o{}rensen}}]{LongDistEntDist_PRL10}
\bibinfo{author}{\bibfnamefont{J.~B.} \bibnamefont{Brask}},
  \bibinfo{author}{\bibfnamefont{I.}~\bibnamefont{Rigas}},
  \bibinfo{author}{\bibfnamefont{E.~S.} \bibnamefont{Polzik}},
  \bibinfo{author}{\bibfnamefont{U.~L.} \bibnamefont{Andersen}},
  \bibnamefont{and} \bibinfo{author}{\bibfnamefont{A.~S.}
  \bibnamefont{S\o{}rensen}}, \bibinfo{journal}{Phys. Rev. Lett.}
  \textbf{\bibinfo{volume}{105}}, \bibinfo{pages}{160501}
  (\bibinfo{year}{2010}).

\bibitem[{\citenamefont{Lim et~al.}(2016)\citenamefont{Lim, Joo, Spiller, and
  Jeong}}]{CV_LossyEntSwappingPRA16}
\bibinfo{author}{\bibfnamefont{Y.}~\bibnamefont{Lim}},
  \bibinfo{author}{\bibfnamefont{J.}~\bibnamefont{Joo}},
  \bibinfo{author}{\bibfnamefont{T.~P.} \bibnamefont{Spiller}},
  \bibnamefont{and} \bibinfo{author}{\bibfnamefont{H.}~\bibnamefont{Jeong}},
  \bibinfo{journal}{Physical Review A} \textbf{\bibinfo{volume}{94}},
  \bibinfo{pages}{062337} (\bibinfo{year}{2016}).



\end{thebibliography}

\end{document}